\documentclass[12pt]{article}

\usepackage{scicite}

\usepackage{times}

\newenvironment{sciabstract}{%
\begin{quote} \bf}
{\end{quote}}

\usepackage[english]{babel}

\usepackage{tikz} 
\usetikzlibrary{fit,positioning}
\usetikzlibrary{matrix}
\usetikzlibrary{calc}
\usetikzlibrary{arrows}
\tikzstyle{int}=[draw, fill=white!8, minimum size=2em]
\tikzstyle{init} = [pin edge={to-,thin,black}]

\usepackage{hyperref}

\usepackage{amsmath}
\usepackage{amssymb}
\usepackage{xspace}
\usepackage{graphicx, subfig}               
\usepackage{setspace}

\usepackage{lineno}

\usepackage{color}

\newcommand{\ac}{\color{red} }  
\newcommand{\ca}{ \color{black}} 
\newcommand{\cutac}[1]{\ac [cut] \ca} 

\newcommand{\ak}{\color{green} }  
\newcommand{\ka}{ \color{black}} 
\newcommand{\cutak}[1]{\ak [cut] \ka} 

\title{
Forecasting hospital demand during COVID-19 pandemic outbreaks
}

\author
{Marcos A. Capistr\'an,$^{1\ast}$ Antonio Capella,$^{2}$ J. Andr\'es Christen$^{1}$\\
\\
\normalsize{$^{1}$Centro de Investigaci\'on en Matem\'aticas, CIMAT-CONACYT,}\\
\normalsize{Jalisco S/N, Valenciana, Guanajuato, GTO, Mexico.}\\
\normalsize{$^{2}$Instituto de Matem\'aticas, UNAM, Circuito Exterior, CU, CDMX, Mexico}\\
\\
\normalsize{$^\ast$To whom correspondence should be addressed; E-mail:  marcos@cimat.mx .}
}

\date{June 2, 2020}

\begin{document}

\baselineskip24pt

\maketitle

\begin{sciabstract}
We present a compartmental SEIRD model aimed at forecasting hospital occupancy in metropolitan areas during the current COVID-19 outbreak.
The model features asymptomatic and symptomatic infections with detailed 
hospital dynamics. We model explicitly  branching probabilities and non exponential residence times in each latent and infected compartments. 
Using both hospital admittance confirmed cases and deaths we infer the contact rate and the initial conditions of the dynamical system, considering break points to model lockdown interventions.
Our Bayesian approach allows us to produce timely probabilistic forecasts of hospital demand.
The model has been used by the federal government of Mexico to assist public policy, and has been applied for the analysis of more than
70 metropolitan areas and the 32 states in the country.
\end{sciabstract}

\section*{Introduction}

The ongoing COVID-19 pandemic has posed a major challenge to 
public health systems of all countries with the imminent risk of saturated hospitals and patients not receiving proper medical care. 
Although the scientific community 
and public health authorities had insight regarding the risks and preparedness measures required in the onset of a zoonotic pandemic, our knowledge of the fatality and spread rates of  COVID-19 remains limited~\cite{ferguson2020impact,verity2020estimates,novel2020epidemiological,zhou2020}. 
In terms of disease handling, two leading issues determining the current situation are the lack of a pharmaceutical treatment and our 
inability to diagnose the asymptomatic infection of 
COVID-19~\cite{gandhiasymptomatic,mizumoto2020diamonprincess,Daym2020asymptomatic}.  

Under the current circumstances, control measures reduce new infections by limiting the number of contacts through mitigation and suppression~\cite{ferguson2020impact}. Mitigation includes social distancing, testing, tracing and isolating of infected individuals, while suppression imposes temporary cancellation of non-essential activities. Most certainly, mitigation and suppression pose a burden on the economy, affecting more those individuals living in low income conditions, thus affecting the capacity of the population as a whole to comply with control measures. 

Undoubtedly, one key task during the early pandemic response efforts
is using epidemiological records and mathematical and statistical modeling to forecast
excess hospital care demand, with formal quantified uncertainty. In this paper we pose a 
compartmental SEIRD model that takes into account both asymptomatic and symptomatic infection,
including hospital dynamics.  We model explicitly the residence time in each latent and infected 
compartments~\cite{champredon2018equivalence,wearing2005erlang_seir_models} and we use
records of daily confirmed cases and deaths to pose a statistical model that accounts for data 
overdispersion~\cite{linden2011using,zarebski2017model}. 
Furthermore, we use Bayesian inference to estimate both the initial state of the governing equations and the contact rate in order to make probabilistic forecasts of the required hospital beds, including the number of intensive care units.  We have applied this model to forecast hospital demand in
metropolitan areas of Mexico. We remark that this model has been used by Mexican federal public
health authorities to assist decision making during the COVID-19 pandemic.

\subsection*{Contributions and limitations}

Broadly speaking, data-driven epidemiological models are built out of necessity of making forecasts. 
There are many lessons learnt on emergency preparedness and epidemic surveillance from 
previous pandemic events: AH1N1 influenza~\cite{cordova20172009}, MERS~\cite{ha2016lesson}, SARS~\cite{emanuel2003lessons}, 
Zika~\cite{hoffman2018delays}, Ebola~\cite{gates2015next}, etcetera. However, surveillance data during a pandemic event
often suffers from serious deficiencies such as incompleteness and backlogs. Another important issue is the design of data acquisition
taking into account geographical granularity~\cite{perra2015modeling}. Epidemic surveillance of COVID-19 is no different since 
there is an unknown percentage of asymptomatic infections, and susceptibility is related to economic vulnerability.

In this paper we demonstrate that it is possible to produce accurate  mid--term (several weeks) probabilistic forecasting of COVID-19 hospital pressure, namely hospital beds and respiratory support or mechanical ventilation demands, using confirmed records of cases at hospital admittance and deaths. 

Our analysis uses the Bayesian approach to inverse problems (Bayesian Uncertainty Quantification) as a natural mathematical tool to investigate the pandemic evolution and provide actionable forecasts to public health controls during the pandemic.

On the other hand, since asymptomatic infection is not fully described so far~\cite{park2020reconciling},
it is impossible to forecast the population fraction that will 
be in contact with the virus by the end of the current outbreak. Likewise, although it is possible to make 
model based analysis of scenarios of lockdown exit strategies, reliability is limited due to the lack of 
information regarding population viral seroprevalence.  Therefore, without serological studies in the open
population after an outbreak, it is not possible to assess the original susceptible population. 
This is a key piece of information to model and assess the possible next pandemic outbreaks, which our model does not address at this point. 

The model introduced here assumes the contact rate remains relatively constant for several weeks. 
Nevertheless, the model accounts for break points due to change of policies in the control
measures, such as school closures~\cite{dehning2020inferring}.

Finally, the model does not account for biases due to behavioral changes~\cite{eksin2019systematic,weitz2020moving}.
In particular, our model may underestimate the decay of outbreaks due to possible granularity of contacts,  superspreading events\cite{liu2020secondary} and other factors.

\subsection*{Related work}

There are many modeling efforts aimed at forecasting hospital occupancy during the ongoing COVID-19 
pandemic~\cite{yamana2020projection,henderson2020,imperial2020,covid2020forecasting,moghadas2020projecting,salje2020estimating}.
Broadly speaking, models are informed with evolving information about COVID-19 cases, clinical description of the patient 
residence time in each compartment, fraction of cases per age group, number of deaths, hospital bed occupancy, etc.

Columbia University metapopulation SEIR model~\cite{yamana2020projection} forecasts are based on assumptions relating an effective contact rate with population density at a metropolitan area and social distancing policies. The Covid Act Now model~\cite{henderson2020} forecasts the replacement number $R_t$ and the fraction of infections requiring hospitalization using the Bayesian paradigm to fit a SEIR model to case, hospitalization, death, and recovery counts. The Imperial College response team mathematical model~\cite{imperial2020} uses an unweigthed ensamble of four models 
to produce forecasts of the number of deaths in the week ahead for each country with active transmission. The IHME model~\cite{covid2020forecasting} combines a mechanistic model of transmission with curve fitting to forecast the number of infections and deaths. Moghadas {\it et al.}~\cite{moghadas2020projecting} pose a mechanistic model parametrized with demographic data to project hospital utilization in the United States during the COVID-19 pandemic.
The main goal of Moghadas {\it et al.} is to estimate hospital pressure throughout.  

Other COVID-19 models have been used to explore exit strategies~\cite{karin2020adaptive,di2020expected}, 
the role of recovered individuals as human shields~\cite{weitz2020intervention}, 
digital contact tracing~\cite{ferretti2020quantifying} and break points in the contact rate to account for changes 
in suppression and mitigation policies~\cite{dehning2020inferring}.

\section*{Methods} 

``Models should not be presented as scientific truth'' \cite{jewell2020mathmodels}. They are tools that 
intend to serve a specific purpose, evaluate or forecast some particular aspects of a 
phenomena under certain conditions and ideally should be developed following the 
processes of predictive science \cite{oden2010computer}. Namely, 
identify the quantities of interest (QoI), verify the computational
and mathematical approximation error, including their implication in the 
estimation of QoI, and calibrate the parameters to adjust the model in light of data to bring it closer to
experimental observation. When considering uncertainty, Bayesian inference
may be used to calibrate some key features of the model given measured data. 
Finally, a validation process must be used to build confidence on the accuracy 
of the QoI predictions of the model. 

\subsubsection*{Dynamical model}

\begin{figure}
\label{fig:model}
\begin{center}
\begin{tikzpicture}[node distance=2.cm,auto,>=latex']
   \node [int] (S) {$S$};
   \node [int] (E) [right of = S] {$E$};   
    \node [int] (IA) [above right of = E] {$I^{A}$};            
    \node [int] (IS) [below right of = E] {$I^{S}$};                   
    \node [int] (H1) [below right of = IS] {$H^{1}$};                
    \node [int] (C) [above right of = IS] {$I^C$};                        
    \node [int] (U1) [below right of = H1] {$U^1$};           
    \node [int] (H2) [above right of = H1] {$H ^{2}$};         
    \node [int] (U2) [right of = U1] {$U^2$};               
    \node [int] (H3) [ right of = H2] {$H^3$}; 
    \node [int] (R) [right=4cm of IA] {$R$};                       
    \node [int] (D) [right of = U2] {$D$}; 

   {\footnotesize
    \path[->] (S) edge node {$\lambda$} (E);
    \path[->] (E) edge node  {$(1-f)\sigma_1$} (IA);
    \path[->] (E) edge node  [below left]{ $f\sigma_1$} (IS);  
    \path[->] (IA) edge node  {$\gamma_1$} (R);        
    \path[->] (IS) edge node  [below right]{  $(1-g)\sigma_2$} (C);            
    \path[->] (C) edge node  { $\gamma_2$} (R);                
    \path[->] (IS) edge node [below left] { $g\sigma_2$} (H1);    
    \path[->] (H1) edge node [below right] {  $(1-h)\sigma_3$} (H2);        
    \path[->] (H2) edge node [above left] {  $\gamma_3$} (R);            
    \path[->] (H1) edge node [below left] { $ h\sigma_3$} (U1);            
    \path[->] (U1) edge node [below] {  $\sigma_4$} (U2);                
    \path[->] (U2) edge node [below] {  $i\mu$} (D);                    
    \path[->] (U2) edge node [right] {  $(1-i)\mu$} (H3);                        
    \path[->] (H3) edge node [left]{ $ \gamma_4$} (R);  
    }                                          
\end{tikzpicture}
\caption{\label{fig:diagram} Schematic diagram of model compartments without Erlang sub-compartments.
For a precise definition of parameters see supplementary material. 
}
\end{center}
\end{figure}
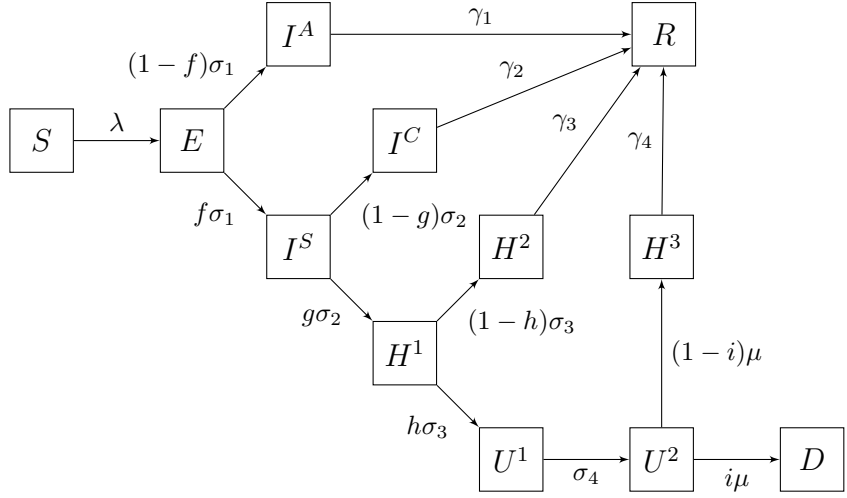

As a proxy of hospital pressure the QoI in our model are: the evolution demand
of no-ICU hospital beds and number of ICU/respiratory support beds. To estimate the 
QoI we developed a full compartmental SEIRD model featuring several compartments
to describe hospital dynamics (see Figure~\ref{fig:diagram} and 
supporting materials, SM) and sub-compartments to describe explicitly residence 
rates as Erlang distributions~\cite{champredon2018equivalence,wearing2005erlang_seir_models}. 
The model has two variants, one with age structure and one that assumes age 
independent dynamics.  Here we describe the only latter (see supporting material 
for some additional comments on the age dependent mode).

Succinctly our model goes as follows (see Figure~\ref{fig:diagram} and SM).
Once the susceptible individuals ($S$) become infected they remain in the incubation compartment 
($E$) for mean time of  $1/\sigma_1$ days (i.e. residence rate $\sigma_1$).  
After the incubation period, exposed individuals become infectious and  
a proportion $f$ become sufficiently sever symptomatic cases ($I^S$) 
to approach a hospital, while the remaining cases become mild to asymptomatic 
($I^A$).  The asymptomatic/mild--symptomatic cases remain infectious  
a mean time of $1/\gamma_1$ days and eventually recover.  

For the symptomatic cases ($I^S$) we assume that after an average time of 
$1/\sigma_2$ days a proportion $g$ of infected individuals will need 
hospitalization ($H^1$), while the rest ($I^C$) will receive ambulatory 
care, recovering after an average convalescent time of  $1/\gamma_2$ 
days in quarantine. 
The hospitalized patients ($H^1$) remain an average time of 
$1/\sigma_3$ days until a fraction $h$ will need assisted 
respiratory measures or ICU care such as mechanical ventilation ($U^1$). 
The remaining fraction $1-h$ of hospitalized patients ($H^2$) will recover 
after an average of $1/\gamma_3$ days. Respiratory-assisted/ICU patients 
($U^1$) remain in that state an average of $1/\sigma_4$ days until a critical 
day is reached when a proportion $i$ of them will die ($D$) and 
the remaining proportion $1-i$ will recover ($H^3$) after an average period 
of $1/\gamma^4$ days.  Similar models have been proposed 
by~\cite{weitz2020intervention,kucharski2020early,ferretti2020quantifying,verity2020estimates}.
For the infection force ($\lambda$) we assume that only  
mild--symptomatic/asymptomatic ($I^A$) and symptomatic ($I^S$) individuals  
spread the infection, that is  
$$
\lambda =\frac{\beta_{A} I^{A} +\beta_{S} I^{S}}{N_{eff}}
$$
where $\beta_A$ and $\beta_S$ are the contact rates of asymptomatic/mild--symptomatic
and symptomatic individuals, respectively.

The model has two kind of parameters that have to be calibrated or inferred,
those related to COVID-19 disease and hospitalization dynamics 
(such as residence times and proportions of individuals that split at each 
bifurcation of the model) and those related to the public response to mitigation
measures such as the contact rates $\beta$'s and the effective number of 
susceptible individuals $N_{eff}$ at the beginning of an outbreak. Some of these
parameters may be estimated from hospital records, be found in recent
literature or inferred from reported cases and deaths, but some 
of remain largely unknown.  In the latter category we have the effective 
number of susceptible individuals $N_{eff}$ at the beginning of an outbreak and 
the fraction $1-f$ of asymptomatic/mild--symptomatic infections.
The number $N_{eff}$ is lower than the full  population for a 
metropolitan area and depends on different aspects, but at least in 
the case of COVID-19 it is likely to be a consequence of unequal 
follow--up of social distancing public policies among the population.  
Reported values of the proportion of asymptomatic/mild--symptomatic 
infections cases $1-f$ range from $10\%$ to $75\%$, and even $95\%$ 
in children population~\cite{mizumoto2020diamonprincess,Daym2020asymptomatic,silal2020estimating}.
Notice that in our model the total number of patients that will visit 
a hospital is given roughly (bounded) by the product $N_{eff}\times f$ and 
the total number of patients admitted to the hospital is given by 
$N_{eff}\times f \times g$, where $g$ is the portion of infected persons 
that need hospitalization. 

We have evidence (see SM) that given 
our choice of observation model (see below), the inference of our QoI 
only depends on the product  $N_{eff}\times f\times g$, and not on 
the value of their individual factors. Since the fraction $g$ is far 
easier to estimate (from hospital  records of admissions and ambulatory
patients) and reported in the literature, we only require to postulate 
a value for the product $N_{eff}\times f$.  

Since our QoI are concerned with hospital pressure, we choose from the 
available data two sources of information for the observational model: 
(A) the registered confirmed COVID-19 patients at hospitals, with 
or without hospitalization, and (B) deceased patients.
Even under an outbreak, these data are reasonable 
consistent and systematic information on the inflow (A, registered patients at hospitals)
and outflow (B, number of hospital deaths), that ``hedge'' the hospital dynamics.
The model assumes constant, and more or less reliable, reported fractions ($g, h, i$) of patients that transit on each hospital dynamics bifurcation.
Intuitively, this provides an explanation for why using only A and B the hospital dynamics (our QoI) may be estimated, while only the product $N_{eff}\times f$ is important and not to a great extent the nominal values of $N_{eff}$ and $f$.

Regarding data availability for our observational model.
Due to administrative reporting delays we discard the last 7 days of reporting. First visit to the hospital date and the registered deceased date are used as time stamps for A and B, respectively. We do not use patient reported symptoms onset as a time stamp.

\subsubsection*{Bayesian Inference}

Undoubtedly, the impact of local testing practices on the number of confirmed cases would need to be analyzed based on the region of interest.
In Mexico testing has been relatively low but consistent.  Patients are tested when arriving to hospitals with probable COVID-19 symptoms and limited testing is done elsewhere; therefore most confirmed COVID-19 cases are limited to A as described above.
Therefore, make our inferences we use both confirm cases (A) and deceased counts (B),
in the sense explained in the previous section.

For inference, we therefore consider daily confirmed cases $c_i$ of patients arriving at $H^1$ and
daily reported deaths $d_i$, for the metropolitan area or region being analyzed.

To account for over dispersion, we use a negative binomial (NB) distribution
$NB( \mu, \omega, \theta)$, following \cite{linden2011using}, with mean $\mu$ and over dispersion parameters $\theta$ and $\omega$ (see SM for details). The use of an over disperse NB has proved its value and adequacy in analyzing data from regions in Mexico (and other regions).
For data $y_i$ we let $y_i \sim NB( p \mu(t_i), \omega, \theta)$, with fixed values for the overdispersion parameters $\omega, \theta$ and an additional reporting probability $p$.
For both confirmed cases and deaths
we found good performance fixing $\omega=2$.
To model daily deaths we fixed $\theta=0.5$ and for daily cases $\theta=1$ implying higher variability for the later.  The reporting probabilities are
0.95 for deaths and 0.85 for cases, with the assumption, as explained above, that the $c_i$'s are confirmed sufficiently severe cases arriving at hospitals.

For daily deaths counts $d_i$ the mean is simply the daily counts $\mu_D(t_i) = D(t_i) - D(t_{i-1})$.  For cases $c_i$, the mean $\mu_c(t_i)$ we consider is the daily flux entering the $H^1$ compartment, which may be calculated as \cite{zarebski2017model}
$$
\mu_c(t_i) = \int_{t_{i-1}}^{t_i} g \sigma_2 I^S_m(t) dt ,
$$
where $I^S_m(t)$ is the last state variable in the $I^S$ Erlang series (see SM).
We calculate the above integral using a simple trapezoidal rule with 10 points (1/10 day).

We assume conditional independence in the data and therefore from the NB model we obtain a likelihood.  Our parameters are the contact rate parameter $\beta$ and crucially we also infer the initial conditions
$E(0), I^A(0), I^S(0)$.  Letting $S(0) = N-(E(0)+I^A(0)+I^S(0))$ and setting the rest of the parameters to zero, we have all initial conditions defined and the model may be solved numerically to obtain $\mu_D$ and $\mu_c$ to evaluate our likelihood.  We use the \textit{lsoda} solver available in the \textit{scipy.integrate.odeint} Python function.

To model interventions, a break point is established at which $\beta = \beta_1$ before and $\beta = \beta_2$ after the intervention day.  This creates a non-linear time dependent $\beta(t)$ \cite{wang2006, dehning2020inferring}.  This additional parameters are then included in the inference.  Normally only the initial $\beta_1$ and an after lockdown (22 March 2020) $\beta_2$ parameters are considered (in Mexico city a third $\beta_3$ was considered for modeling a further local intervention in early April).

We use $Gamma(1,10)$ priors for each $E(0), I^A(0), I^S(0)$, modeling the low, near to 10, and close to zero counts for the number of infectious initial conditions.  The prior for the $\beta$'s are the same, and a priori independent, long tails, log Normal with $\sigma^2 = 1$ and scale parameter 1; that is $log(\beta) \sim N(0,1)$.
To sample from the posterior we resort to MCMC using the ``t-walk'' generic sampler \cite{christen2010twalk}.  The MCMC runs semi automatic, with fairly consistent performances in most data sets.  For any state variable $V$, the MCMC allows us to sample from the posterior predictive distribution for $V(t_i)$, and by plotting sequentially some of its quantiles we may produce predictions with quantified probabilistic uncertainty, as seen in Figure~\ref{fig:cdmx}.

As in the case of climate forecasting, due to the stochastic nature of an pandemic outbreak point-wise estimates such as the maximum a posteriori estimate (MAP) 
do not give good descriptions of the outbreak evolution. No single trajectory of the SEIRD 
model provide a good description of the outbreak evolution, nor give accurate forecasts.



 


\section*{Results}

Local transmission started in a different date in each Mexican metropolitan area provided each one of them has different 
communicability with Mexico City and with the rest of the world. 
On the other hand, all metropolitan areas in the country 
are in lock-down since May 22, with a change of policy due to start at the beginning of the month of June. 

\begin{figure}[!ht]
\begin{center}
\begin{tabular}{c c}
\includegraphics[scale=0.4]{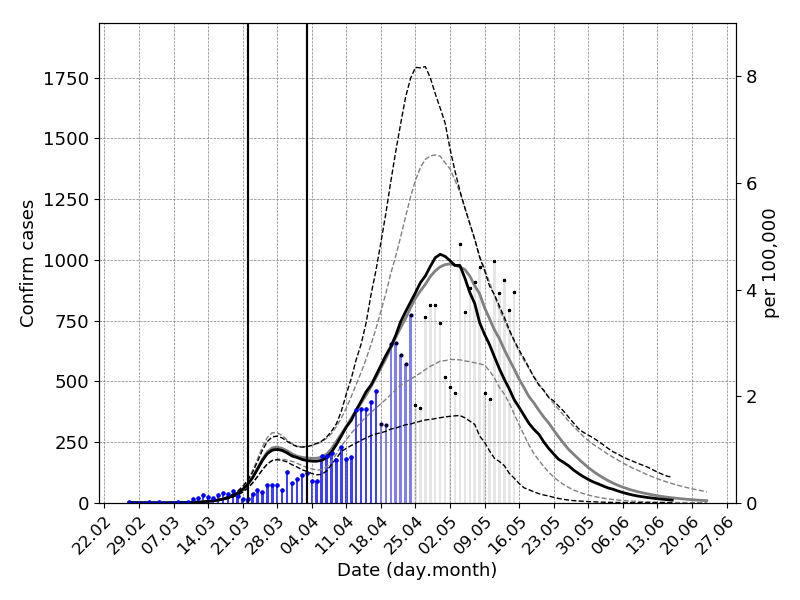}    &
\includegraphics[scale=0.4]{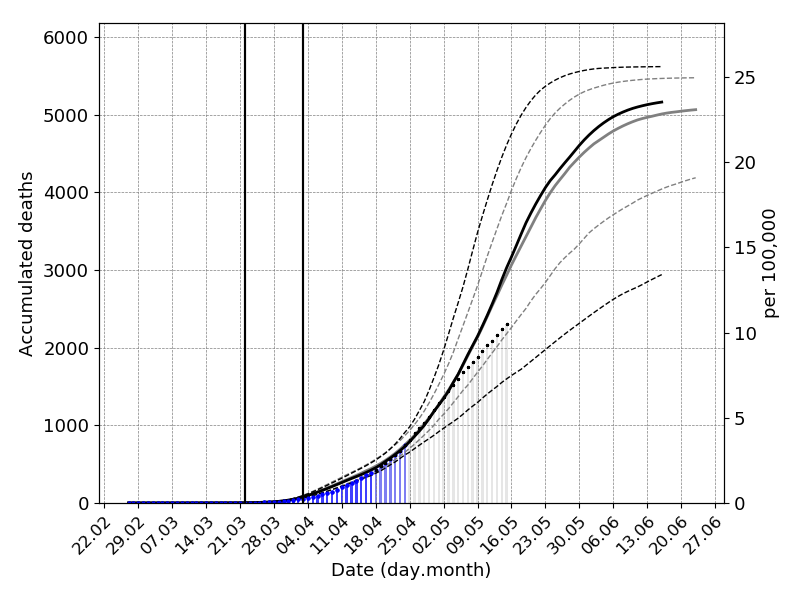} \\
 (a) & (b) \\
\includegraphics[scale=0.4]{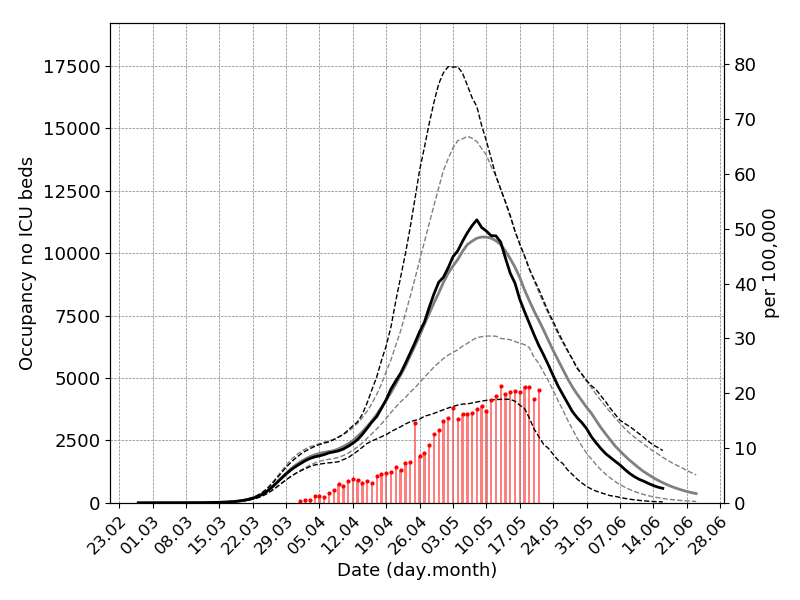}    &
\includegraphics[scale=0.4]{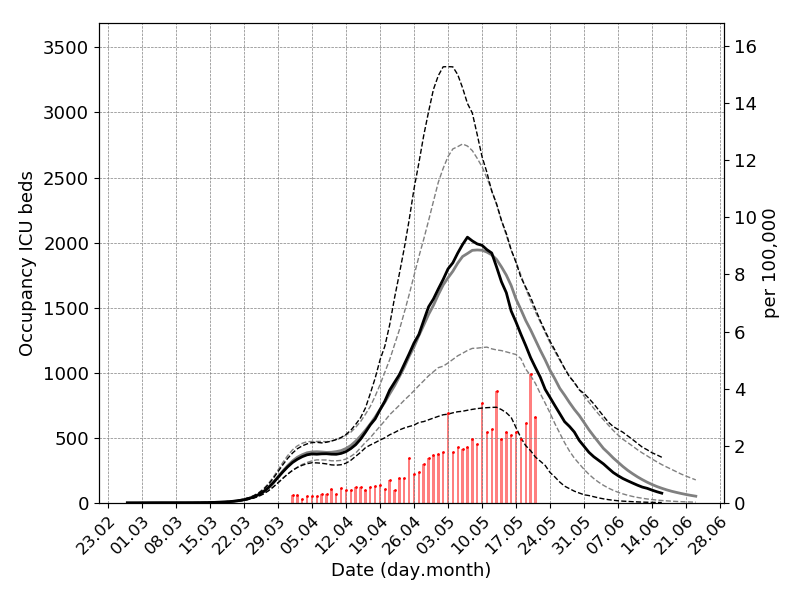} \\
 (c) & (d) \\
\end{tabular}
\end{center}
\caption{Outbreak analysis for Mexico city metropolitan area, using data from April 17 and April 24.  (a) Confirmed cases (b) Deaths (c) No ICU (d) ICU occupied hospital beds. Total population $21,942,666$ inhabitants}
\label{fig:cdmx}
\end{figure}

Figure~\ref{fig:cdmx} (a) shows the model forecast, with quantified uncertainty, of the daily records of COVID-19 
confirmed cases in Mexico City with data trimming to April 17 and April 24 for comparison. 
Figures shows both, the absolute number of cases and the number of cases per 100,000 inhabitants.
Of note, the update in the median, 
shown with a continuous black line, is almost negligible while the update in quantiles 10\% and 90\%,
shown with dotted lines, exhibits a contraction around the median as more data is included. Figure~\ref{fig:cdmx} (b) 
depicts records and forecasts for the aggregated number of deaths. In Figure~\ref{fig:cdmx} (c) and (d) we compare the model
forecasts with hospital bed and ICU occupancy obtained from a secondary official source of epidemiological surveillance.
We remark that the total number of hospital beds and ICU units is consistently overestimated and shifted 
to latter times. Our modeling strategy
was based on daily demand of hospital beds and intensive care units records from {\it Instituto Mexicano del Seguro Social} 
or Mexican Social Security Institute (IMSS) to calibrate the residence time within the hospital. We assumed that it was
necessary to forecast a higher demand in both types of hospital occupancy given the emerging nature of the virus and evidence
of excess of hospital pressure in places like Spain, Italy and New York.

In the supplementary material we show the outbreak analysis for the cities of Canc\'un, Tijuana and La Paz. The rationale for 
choosing these Mexican cities is as follows. Mexico City is among the top ten most populated metropolitan areas in the world and 
comprises roughly one sixth of the Mexican population. Canc\'un and Tijuana are medium size cities with considerable international communicability where the first ones with an outbreak among cities in Mexico.
La Paz is the capital of the Mexican state of Baja California Sur in the Baja California peninsula where the outbreak was relatively small. 
%
Despite the varying features among these cities we forecast the acme of the developing outbreak with up to three weeks for small 
and medium size cities. 


\subsection*{Mexico city, metropolitan area}

\section*{Discussion}

We presented a Bayesian approach to inform a compartmental SEIRD model to produce probabilistic forecasts of hospital pressure during COVID-19 outbreaks.

Besides the examples presented in this paper, 
our model has been applied to other 70 metropolitan 
areas and the 32 states in Mexico
(``ama'' model; \url{https://coronavirus.conacyt.mx/proyectos/ama.html}, in Spanish).
From simple model forecast comparisons it was apparent from early April that pandemic outbreaks ran at different speeds throughout metropolitan areas in the country,
suggesting more regionalized policies than a single nationwide strategy.
Consequently, future re-opening strategies may need to be thought on a regional basis also.

The age independent model has proven to be adequate
to produce accurate forecast for the hospitalization dynamics during current outbreaks. Therefore the added complexity of the age-structure model may not justify its used at this point. However we continue to work in our age structure model. 

Regarding interventions, we model lock-down and social distancing policies by setting break points in 
the contact rates at specific dates and inferring their values. 
This method does not serve to detect change points but it does measure
the effectiveness of these policies and its consequences on our forecast.
Moreover, it makes our model more flexible adapting inference for changing circumstances. 

Our observational model is designed to integrate data after the 
non-linear term in the dynamic model and the epidemic curves we 
inferred and forecast are for the hospitalization dynamics.
Assuming (as the model does) that the rest of the dynamics is 
proportional to these curves, the dynamics of the rest of the 
model will follow --bounded by residence times-- the same paths 
of the hospitalization dynamics. Thereby, our estimates and 
forecasts on hospitalization pressure can be used  as a proxy 
of the full outbreak. In turn we are also able to produce 
probabilistic forecasts for the date when an outbreak will reach its peak. 

As already mentioned, as yet the asymptomatic infection has not been fully described~\cite{park2020reconciling}. 
The confounding effect 
of the effective population that participates in the outbreak 
and the fraction of asymptomatic/mild--symptomatic infections $1-f$
makes it impossible to reliably forecast the population fraction that 
will be in contact with the virus (i.e. the final population viral seroprevalence) at the end of an outbreak.
Likewise, although it is possible to make model 
based analyses of scenarios of lock-down exit strategies, 
scenario reliability estimation is limited due to the lack of information regarding population viral seroprevalence.  Therefore, without 
serological studies in the open population after an outbreak 
it is not possible to asses the final outbreak size. 
This is a key piece of information to model and assess possible 
next pandemic outbreaks, from the possibility of reopening and relaxation of social distancing measures.  Our model as it stands cannot predict multiple outbreaks, but once epidemic data is available, estimation of the previous and new effective population may provide a toll to predict
second outbreaks (for a fixed proxy $f$ probability; work in progress).

There has been substantial discussion regarding testing 
strategies for prevalence of COVID-19 cases in different countries.
These strategies may serve dissimilar purposes and it should 
be clear the intended use of each strategy. Due to the lack of 
description of the asymptomatic infection most of 
the strategies for large amount of testing that includes 
the asymptomatic infections will produce biased data and
these biases are hard to identify. Mexico is one of the
countries with the smallest ratio of COVID-19 testing~\cite{}.
However, our results show that it is possible to produce reliable forecast for hospital pressure during an outbreak with this relative low amount of testing exclusively at hospital admissions (and deceased). The advantage of this testing strategy is that it is systematic, biases are well identified and it is possible to develop a model well adapted to this observational method.

\section*{Acknowledgments}

The authors wish to thank Elena \'Alvarez-Buylla, Paola Villarreal (CONACYT) and Hugo L\'opez-Gatell (Secretar\'ia de Salud) for their support and comments for improving this forecast model and
also Instituto Mexicano de Seguridad Social (IMSS) for sharing data to adjust the hospitalization dynamics.
Very special thanks to CIMAT technicians
Judith Esquivel-V\'azquez and Oscar Gonz\'alez-V\'azquez (CIMAT-CONACYT), help in retrieving data and
running our programs at a CIMAT cluster.  We also thank the CONACYT COVID-19 response group, for
additional comments in the developing of our model.
The authors are partially founded by CONACYT CB-2016-01-284451 grant.
AC was partially supported by UNAM PAPPIT–IN106118 grant.

\bibliographystyle{Science}

\newpage

\section*{Supplementary materials}




\section*{Other examples}
\subsection*{Tijuana, Mex}

\begin{figure}[!ht]
\begin{center}
\begin{tabular}{c c}
\includegraphics[scale=0.4]{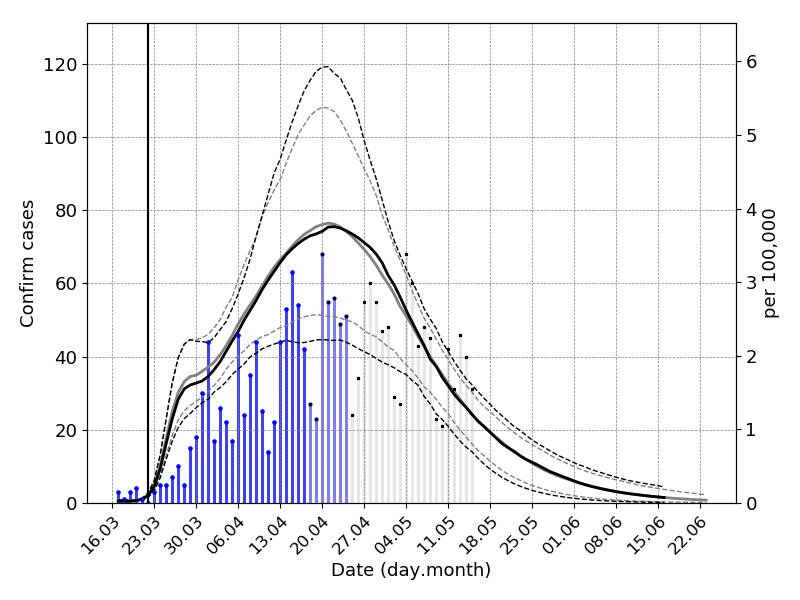}    &
\includegraphics[scale=0.4]{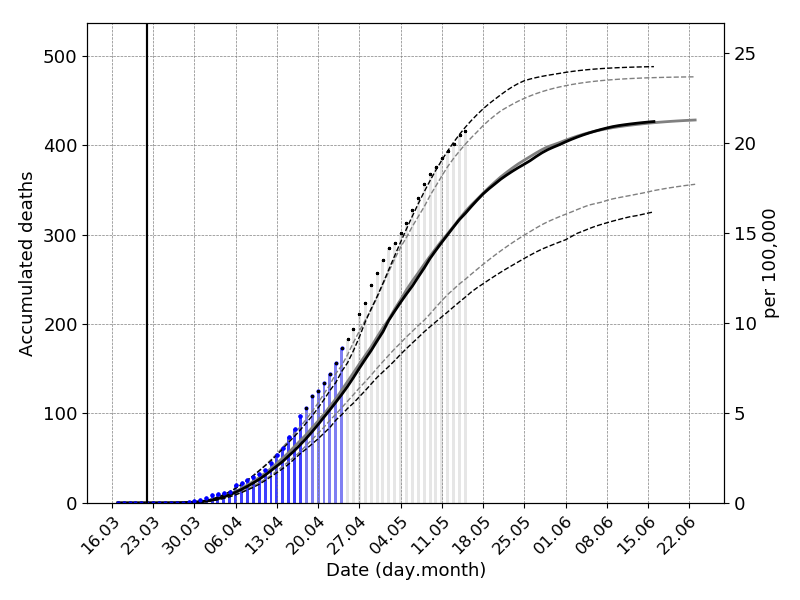} \\
 (a) & (b) \\
\includegraphics[scale=0.4]{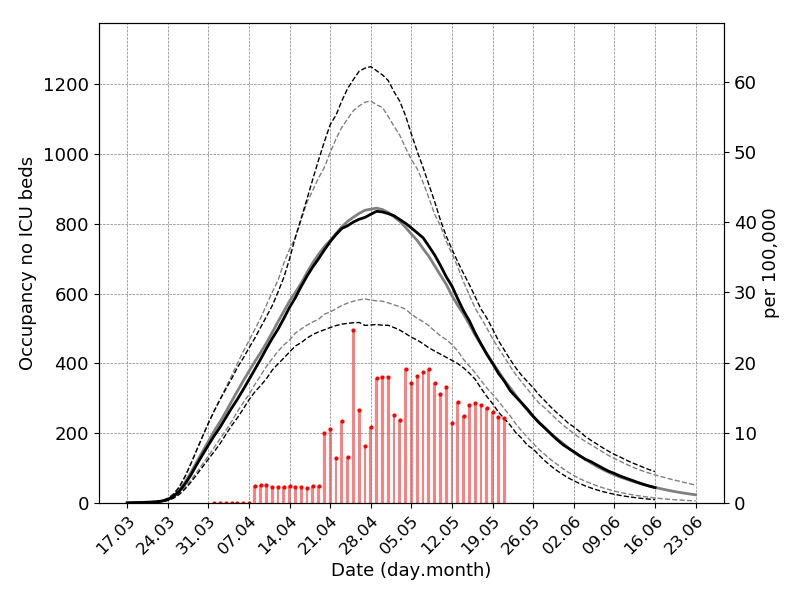}    &
\includegraphics[scale=0.4]{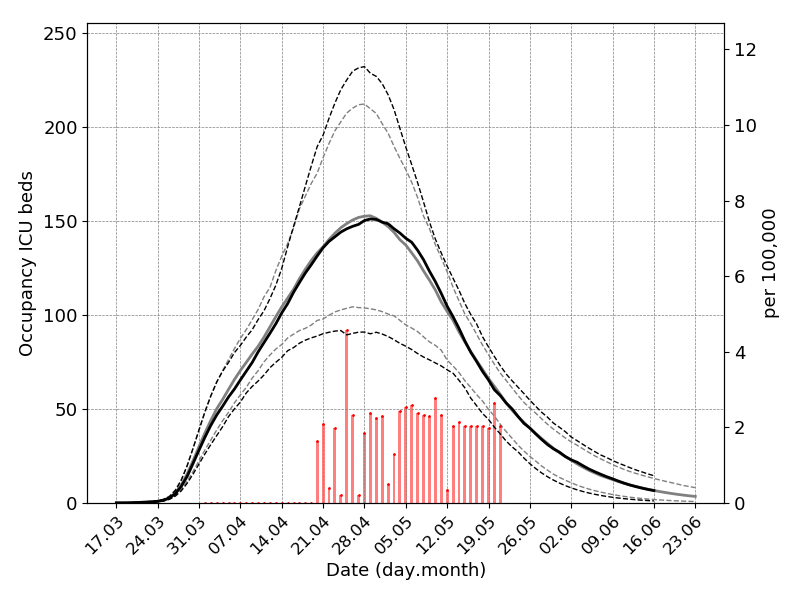} \\
 (c) & (d) \\
\end{tabular}
\end{center}
\caption{\label{fig:tijuana} Outbreak analysis for Tijuana metropolitan area, using data from 17 April and 24 April. (a) Confirmed cases (b) Deaths (c) No ICU (d) ICU occupied hospital beds. Total population $2,011,247$ inhabitants}
\end{figure}

\clearpage

\subsection*{Cancun, Mex}

\begin{figure}[!ht]
\begin{center}
\begin{tabular}{c c}
\includegraphics[scale=0.4]{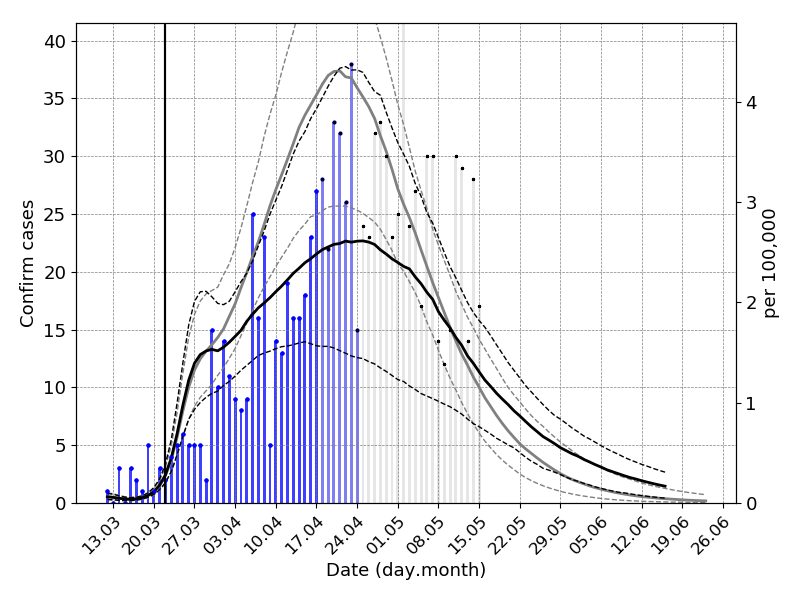}    &
\includegraphics[scale=0.4]{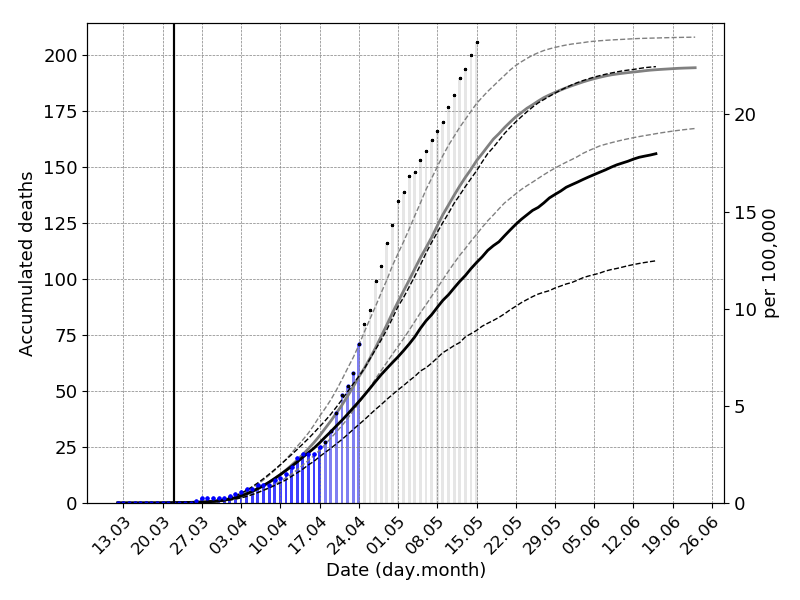} \\
 (a) & (b) \\
\includegraphics[scale=0.4]{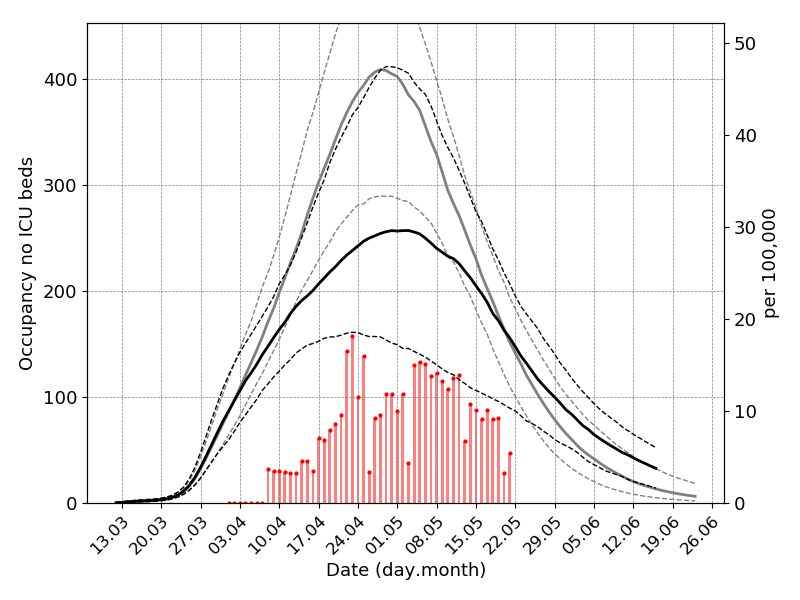}    &
\includegraphics[scale=0.4]{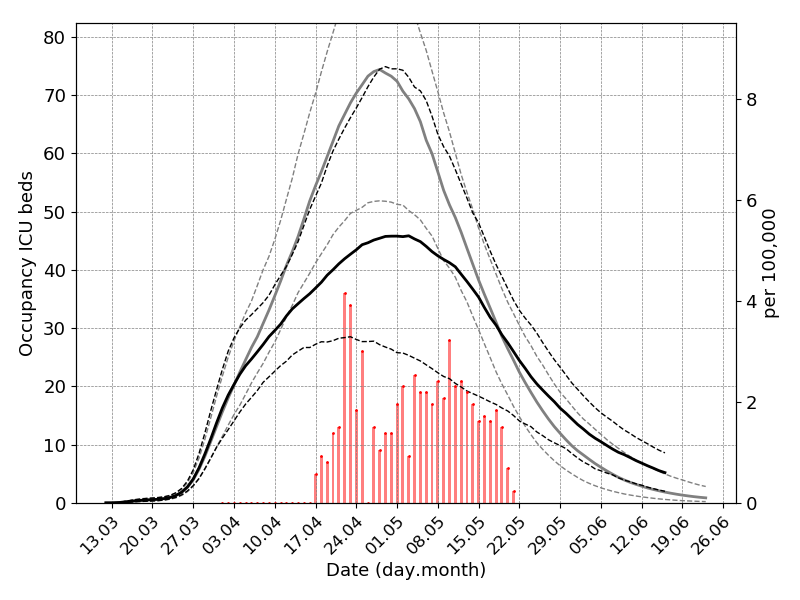} \\
 (c) & (d) \\
\end{tabular}
\end{center}
\caption{\label{fig:cancun} Outbreak analysis for Cancun metropolitan area, using data from 17 April and 24 April. (a) Confirmed cases (b) Deaths (c) No ICU (d) ICU occupied hospital beds. Total population $867,768$ inhabitants.}
\end{figure}

\clearpage

\subsection*{La Paz}

In cases where there are small counts of confirmed 
cases and deaths we included a fictitious intervention 
point after the first 10 confirmed cases~\cite{cori2013Rt}. In terms of 
the outbreak evolution this serves the purpose of 
distinguish the imported cases from the community 
spread where the model should be applied. 

\begin{figure}[!ht]
\begin{center}
\begin{tabular}{c c}
\includegraphics[scale=0.4]{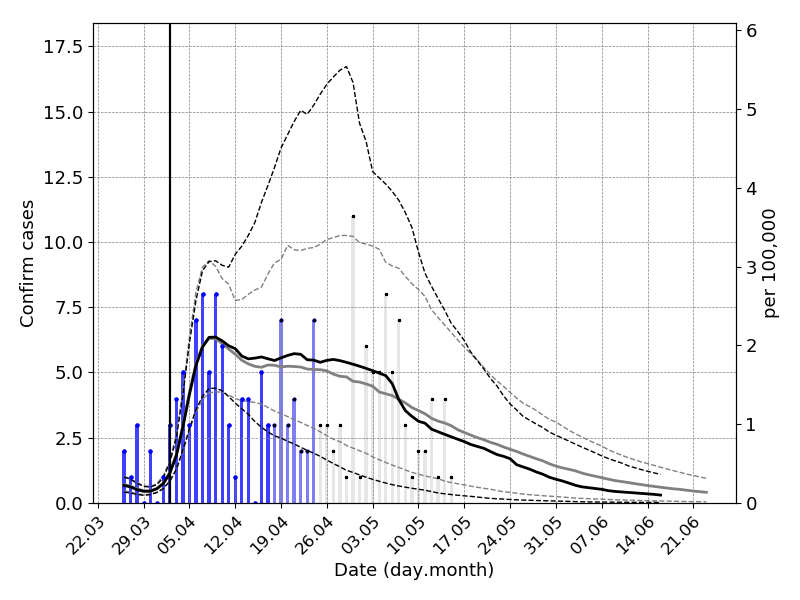}    &
\includegraphics[scale=0.4]{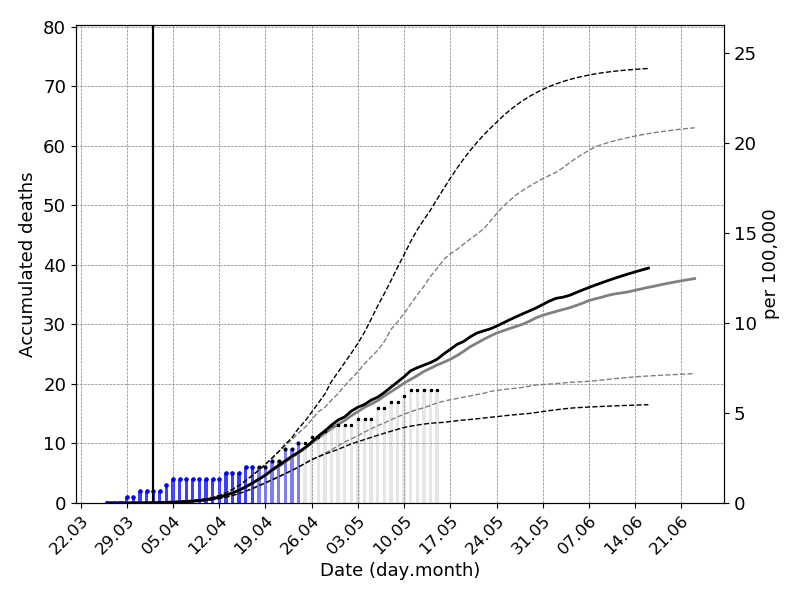} \\
 (a) & (b) \\
\includegraphics[scale=0.4]{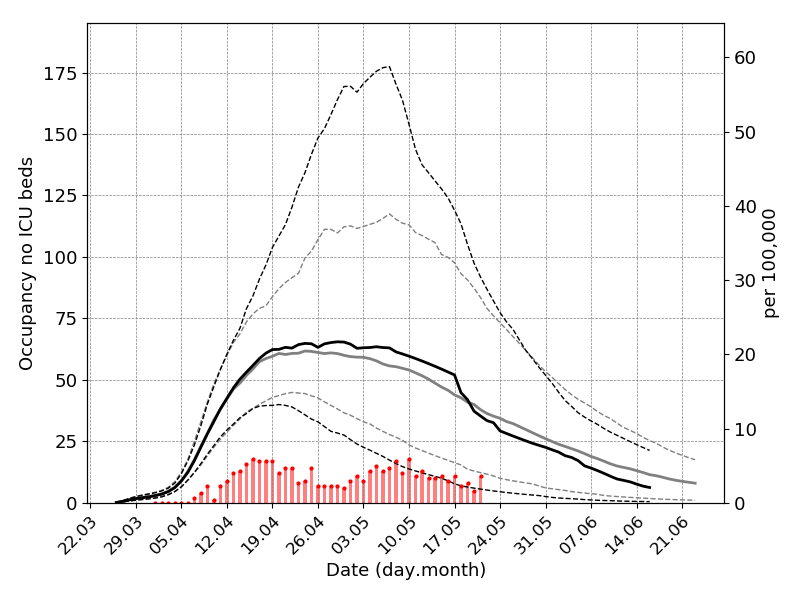}    &
\includegraphics[scale=0.4]{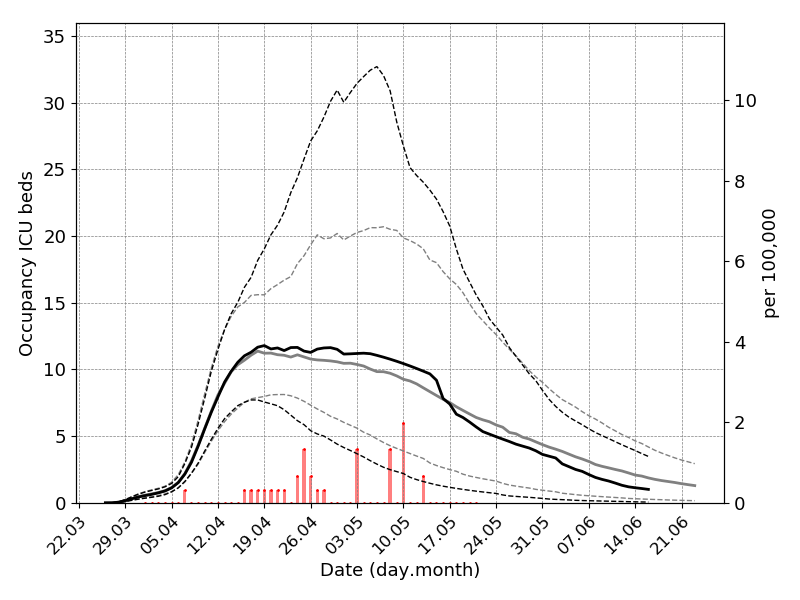} \\
 (c) & (d) \\
\end{tabular}
\end{center}
\caption{\label{fig:lapaz} Outbreak analysis for La Paz metropolitan area, using data from 17 April and 24 April. (a) Confirmed cases (b) Deaths (c) No ICU (d) ICU occupied hospital beds.  Total population $301,961$ inhabitants.}
\end{figure}

\subsection*{Data sources}
Delimitation of the metropolitan areas and population was taken from \url{https://www.gob.mx/conapo/documentos/delimitacion-de-las-zonas-metropolitanas-de-mexico-2015.html}  and  \url{https://datos.gob.mx/busca/dataset/proyecciones-de-la-poblacion-de-mexico-y-de-las-entidades-federativas-2016-2050}, respectively. The official records of COVID-19 confirmed cases and dead 
where taken from  https://www.gob.mx/salud/documentos/datos-abiertos-152127

\section*{Model}

We developed a dynamic transmission compartmental model to simulate the 
spread of the novel coronavirus SARS-CoV-2.  A description of the state variables is found in
Table~\ref{tab:state_vars}.  Additionally and ``Erlang series'' is included for most of these state
variables to account for non-exponential residente times; see below.  The model may be described conceptually with the graph in Figure~\ref{fig:diagram}.
Without showing the Erlang series for the sub-compartments the system
of equations in the model is as follows:
\begin{align*}
\frac{dS}{dt} & =  
-\frac{\left(\beta_{A} I^{A} +\beta_{S} I^{S} \right)}{N_{eff}} S ~~ &  \frac{dE}{dt} & =  \frac{\left(\beta_{A} I^{A} +\beta_{S} I^{S} \right)}{N_{eff}}  S - \sigma_1 E\\
\frac{dI^A}{dt} & =   (1-f) \sigma_1 E - \gamma_1 I^A ~~ & \frac{dI^S}{dt} & =   f \sigma_1 E - \sigma_2 I^S\\
\frac{dI^C}{dt} & =  (1-g) \sigma_2 I^S - \gamma_2 I^C 
~~ & \frac{dH^1}{dt} & = g \sigma_2 I^S - \sigma_3 H^1  \\
\frac{dH^2}{dt} & = (1-h)\sigma_3 H^1 - \gamma_3 H^2  
~~ &  \frac{dU^1}{dt} & =  h \sigma_3 H^1 - \sigma_4 U^1\\
\frac{dU^2}{dt} & = \sigma_4 U^1 - \mu U^2 
~~ &  \frac{dD}{dt} & = i \mu U^2 \\
\frac{dH^3}{dt} & = (1-i)\mu U^2 - \gamma_4 H^3
~~ &  \\
\frac{dR}{dt} & = \gamma_1 I^A+\gamma_2 I^C+\gamma_3 H^2 +\gamma_4 H^3 ~~. &  
\end{align*}
A brief description of all the parameters in our model may be found in Table~\ref{tab:parameters}.

\begin{table} 
\begin{caption}
{\label{tab:state_vars} Description of the state variables for our dynamic transmission model}
\end{caption}
\begin{center}
    \begin{tabular}{ c  l }
    {\bf Variable} & {\bf Description} \\ 
    \hline
   
    $S$ 	& Susceptibles \\
    $E$ 	& Latent individuals  \\
   $I^{A}$ 	& Asymptomatic/mild--symptomatic individuals   \\ 
   $I^{S}$ 	& Symptomatic individuals \\       
   $I^C$ 	& Out-patients \\
   $H^1$ 	& Hospitalized patients, initial stage   \\              
   $H^2$ 	& Hospitalized patients (no ICU)    \\               
   $U^1$ 	& Hospitalized patients (ICU or respiratory support) \\              
   $U^2$ 	& Hospitalized patients (ICU or respiratory support) critical day   \\
   $H^{3}$  & Hospitalized patients recovering after ICU or respiratory support \\   
   $R$ 	    & Recovered   \\
   $D$ 	    & Deceased \\     
     \hline           
    \end{tabular}
\end{center}    
\end{table}

\begin{figure}
\begin{center}
\begin{tikzpicture}[node distance=2.cm,auto,>=latex']
   \node [int] (S) {$S$};
   \node [int] (E) [right of = S] {$E$};   
    \node [int] (IA) [above right of = E] {$I^{A}$};            
    \node [int] (IS) [below right of = E] {$I^{S}$};                   
    \node [int] (H1) [below right of = IS] {$H^{1}$};                
    \node [int] (C) [above right of = IS] {$I^C$};                        
    \node [int] (U1) [below right of = H1] {$U^1$};           
    \node [int] (H2) [above right of = H1] {$H ^{2}$};         
    \node [int] (U2) [right of = U1] {$U^2$};               
    \node [int] (H3) [ right of = H2] {$H^3$}; 
    \node [int] (R) [right=4cm of IA] {$R$};                       
    \node [int] (D) [right of = U2] {$D$}; 

   {\footnotesize
    \path[->] (S) edge node {$\lambda$} (E);
    \path[->] (E) edge node  {$(1-f)\sigma_1$} (IA);
    \path[->] (E) edge node  [below left]{ $f\sigma_1$} (IS);  
    \path[->] (IA) edge node  {$\gamma_1$} (R);        
    \path[->] (IS) edge node  [below right]{  $(1-g)\sigma_2$} (C);            
    \path[->] (C) edge node  { $\gamma_2$} (R);                
    \path[->] (IS) edge node [below left] { $g\sigma_2$} (H1);    
    \path[->] (H1) edge node [below right] {  $(1-h)\sigma_3$} (H2);        
    \path[->] (H2) edge node [above left] {  $\gamma_3$} (R);            
    \path[->] (H1) edge node [below left] { $ h\sigma_3$} (U1);            
    \path[->] (U1) edge node [below] {  $\sigma_4$} (U2);                
    \path[->] (U2) edge node [below] {  $i\mu$} (D);                    
    \path[->] (U2) edge node [right] {  $(1-i)\mu$} (H3);                        
    \path[->] (H3) edge node [left]{ $ \gamma_4$} (R);  
    }                                          
\end{tikzpicture}
\caption{\label{fig:graph} Schematic diagram of model compartments without Erlang sub-compartments.}
\end{center}
\end{figure}
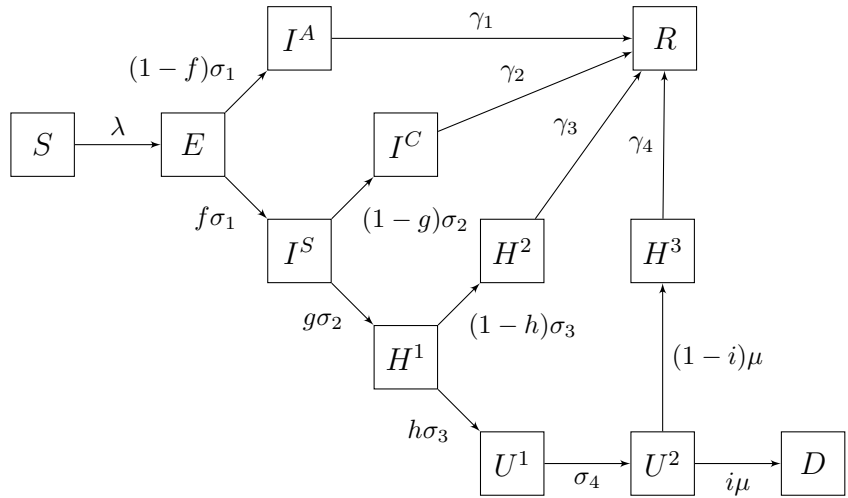

\begin{table} 
\begin{caption}
{\label{tab:parameters} Description of the parameters included in our model.}
\end{caption}
{\small
\begin{center}
    \begin{tabular}{ c  l c}
    {\bf Parameter} & {\bf Description} & {\bf Units}\\ 
    \hline
   $N_{eff}$      & Number of initially susceptible individuals in the population  & -- \\	
   $\beta_A$  & Transmission rate of asymptomatic/mild--symptomatic individuals (asx/mild--sym) & per day  \\
   $\beta_S$  & Transmission rate of symptomatic individuals & per day \\
   $ \kappa $ & Relative strength between the transmission the rate of asx/mild--sym and symptomatic   \\
    $f$  	  & Portion of infected persons with strong enough symptoms to visit a hospital  \\
   $g$ 	& Portion of infected persons that need hospitalization    \\ 
   $h$ 	& Portion of hospitalized patients requiring respiratory support or ICU care     \\       
   $i$ 	& Portion of Respiratory-assisted or ICU patients deceased\\
   $1/\sigma_1$ 	& Average incubation time  & day \\              
   $1/\sigma_2$ 	& Average time from symptomatic onset to hospital visit  & day \\          
   $1/\sigma_3$ 	& Average time from hospital admission to respiratory support or ICU care & day \\              
   $1/\sigma_4$ 	& Average time with respiratory support or ICU care & day \\
   $1/\mu$    & Average length of critical stage of respiratory--support/ICU between death and recovery & day \\
   $1/\gamma_1$  & Average time that asymptomatic/mild--symptomatic individuals remain
   infectious & day\\
   $1/\gamma_2$ 	& Average time of symptomatic individuals that recover without visiting a hospital & day \\
   $1/\gamma_3$     & Average time from hospital admission to hospital discharge & day \\ $1/\gamma_4$     & Average time from respiratory--support/ICU care release  to hospital discharge & day\\ 
     \hline           
    \end{tabular}
\end{center}    
}
\end{table}

\subsubsection*{Infection force and basic reproductive number $\mathcal{R}_0$}

For the infection force ($\lambda$) we assume that the only individuals 
that spread the infection correspond to the mild--symptomatic/asymptomatic
($I^A$) and symptomatic individuals ($I^S$) before their contact with the
health care system or doctor, e.g.
$$
\lambda =\frac{\beta_{A}I^{A}+\beta_{S}I^{S}}{N_{eff}}
$$

We compute the basic reproductive number $R_0$ of the epidemic by the 
next generation matrix method~\cite{van2002reproduction} and obtain  
$$
R_0 = (1-f)\frac{\beta_A}{\gamma_1}+f\frac{\beta_S}{\sigma_2}.
$$

\subsubsection*{Values of model parameters}

Since our QoI are related with the hospital pressure we choose 
all parameters conservatively. For each metropolitan area we assume that $N_{eff}$ corresponds to its full population as 
defined by Instituto Nacional de Estadistica y Geografia (INEGI). The values of the transition probabilities are summarized in Table~\ref{tab:trnsprob}.

\begin{table}[h!] 
\begin{caption}
{\label{tab:trnsprob} Transition probabilities at bifurcations in the model}
\end{caption}
{\small
\begin{center}
    \begin{tabular}{ c  l c}
    {\bf Parameter} & {\bf Value} & {\bf Reference}\\ 
    \hline
    $f$  & 0.05 & postulated \\
   $g$ 	&  0.04375 & \cite{ferguson2020impact}, IMSS    \\ 
   $h$ 	&   0.25 & \cite{ferguson2020impact}, IMSS\\      
   $i$ 	& 0.5 & \cite{ferguson2020impact}, IMSS\\
     \hline           
    \end{tabular}
\end{center}    
}
\end{table}

\subsubsection*{Erlang series and sub-compartments}

In order to make the intrinsic generation-interval of the renewal equation 
in each compartment more realistic we divide each compartment of the model 
into $m$ equal sub--compartments to generate an Erlang--distributed waiting 
time \cite{champredon2018equivalence}. The Erlang distributions of each 
compartment is calibrated by two  parameters: the rate $\lambda_E$ and 
the shape $m$, a positive integer that corresponds to the number of 
sub--compartments on the model. In terms of these parameters the mean 
of the Erlang distribution is $m/\lambda_E$, this mean correspond to 
the average times in the dynamic model. 

We use recent publications and information generously shared by the 
Instituto Mexicano de Seguridad Social (IMSS) to estimate the average 
time and the shape parameter of the Erlang series in each compartment.
Details of Erlang series lengths, residence times and the imputed values
may be found in Table~\ref{tab:erlang}.

\begin{table}[h!]
\caption{\label{tab:erlang} Average times and Erlang shape parameters for sub-compartments}
{
\begin{center}
    \begin{tabular}{ c  c  c  c  c }
    {\bf Variable}  & {\bf Rates}& {\bf Average time} & {\bf Erlang shape $m$}  & {\bf Reference} \\  \hline
    $S$ 	 	& $\beta_S$ 		& Inferred & 1 & -- \\   
   $E$ 	 	& $1/\sigma_{1}$	 & 5 days  & 4  &\cite{verity2020estimates}\\ 
   $I^{A}$ 	  & $1/\gamma_{1}$	& 7 days & 3 & \cite{eurosurveillance2020updated}\\    
   $I^{S}$ 	  & $1/\sigma_{2}$	& 4 days  & 3 & \cite{zhang2020evolving}\\       
   $I^C$ 	 & $1/\gamma_{1}$	& 7 days & 3 & \cite{buchholz2020modellierung} \\ 
   $H^1$ 	  & $1/\sigma_{3}$	& 2 days & 10 &\cite{novel2020epidemiological} \\ 
   $H^2$ 	  & $1/\gamma_{3}$	& 10 days & 3 & \cite{novel2020epidemiological}\\
   $U^1$ 	 & $1/\sigma_{4}$	&  10 days & 3 &  IMSS\\                
   $U^2$ 	  & $1/\mu$	& 1 day & 1 &\cite{buchholz2020modellierung}\\                      
   $H^{3}$   & $1/\gamma_{4}$	& 4 days & 5 & \cite{novel2020epidemiological} \\          
   $R$ 	  & None	& -- & -- & -- \\ 
   $D$  & None	& -- & -- & --\\                         \hline                         
    \end{tabular}
\end{center}
}
\end{table}

\subsubsection*{Relative strength between the transmission the rate of asymptomatic/mild--symptomatic and symptomatic}

In our methodology we aim to infer the force of the infection $\lambda$. 
This  parameter is defined in terms of contact rate of 
asymptomatic/mild–symptomatic  individuals $\beta_A$ and contact 
rate of symptomatic individuals $\beta_S$. Due to the functional dependence 
of $\lambda$ in these parameters there is a lack of identifiability between 
$\beta_A$ and $\beta_S$ that can not be resolved without further assumptions. 
We assume that the relative strength between the transmission 
rate of asymptomatic/mild--symptomatic and symptomatic is modelled as
a fixed ratio $\kappa$. We model the value of $\kappa$ directly as the ratio
of the viral load of symptomatic and asymptomatic/mild--symptomatic patients
\cite{zou2020sars,he2020temporal} and is fixed to $\kappa = 2$. Hence, the force of infection becomes $\lambda = \beta^S(I^S+\kappa I^A) /N_{eff}$.

%

\subsection*{Data and observational model}

To make our inferences we use both confirm cases and deceased counts.
In some regions sub reporting of COVID-19 related deaths may become relevant, specially in places hit by a severe outbreak \cite{MMWR2020}.  Nonetheless, deaths are a more reliable data source to estimate a COVID-19 outbreak, specially for the purposes of hospital demand.  Certainly, daily confirm cases are also needed for more accurate predictions as a proxy for outbreak dynamics.  The problem here is that the number of confirmed cases depends heavily on local practices, in particular in relation to the intensity of testing, adding a level of complication if testing intensity has varied due to ambiguous policies.
Undoubtedly, the impact of local testing practices on the number of confirmed cases would need to be analyzed based on the region of interest.  In Mexico testing has been relatively low but consistent.  Patients are tested when arriving to hospitals with probable COVID-19 symptoms and limited testing is done elsewhere.  For inference, we therefore consider daily counts of patients arriving at $H^1$, as we explain next.

Accordingly, as explained in the main text, we use daily reported deaths $d_i$ and daily confirmed cases $c_i$, for the metropolitan area or region being analyzed, to perform our inferences.  The first default model for count data is a Poisson distribution, however, epidemiological data tends to be over disperse and an over disperse generalized Poisson distribution may be needed to correctly, and safely, model these type of data, as it is the case in other ecological studies. \cite{linden2011using} suggest the use of and over disperse negative binomial (NB) model.  We have followed their suggestion, and the use of an over disperse NB has proved its value and adequacy in analyzing data from regions in Mexico, and elsewhere, as we have already illustrated.

Following \cite{linden2011using} the NB distribution is re parametrized in terms of its mean $\mu$ and ``overdispersion'' parameters $\theta$ and $\omega$, with $r = \frac{\mu}{\omega - 1 + \theta \mu}$ and $p = \frac{1}{\omega + \theta \mu}$, the number of failures before stopping and the success probability, respectively, in the usual NB parametrization. For data $y_i$ we let $y_i \sim NB( p \mu(t_i), \omega, \theta)$, with fixed values for the overdispersion parameters $\omega, \theta$ and an additional reporting probability $p$.  The index of dispersion is $\sigma^2/\mu = \omega + \theta \mu$.  Over dispersion with respect to the Poisson distribution is achieved when $\omega > 1$ and the index of dispersion increases with size if $\theta \neq 0$; both desirable characteristics in outbreak data, adding variability as counts increase.  In both cases we found good performance fixing $\omega=2$.
To model daily deaths we fixed $\theta=0.5$ and for daily cases $\theta=1$ implying higher variability for the later.  The reporting probabilities are
0.95 for deaths and 0.85 for cases, with the assumption, as explained above, that the $c_i$'s are confirmed sufficiently severe cases arriving at hospitals.

For daily deaths counts $d_i$ the mean is simply the daily counts $\mu_D(t_i) = D(t_i) - D(t_{i-1})$.  For cases $c_i$, the mean $\mu_c(t_i)$ we consider the daily flux entering the $H^1$ compartment, which may be calculated as \cite{zarebski2017model}
$$
\mu_c(t_i) = \int_{t_{i-1}}^{t_i} g \sigma_2 I^S_m(t) dt ,
$$
where $I^S_m(t)$ is the last state variable in the $I^S$ Erlang series.  We calculate the above integral using a simple trapezoidal rule with 10 points (1/10 day).

\subsection*{Modeling interventions and Bayesian Inference}

We assume conditional independence in the data and therefore from the NB model we obtain a likelihood.  Our parameters are the contact rate parameter $\beta$ and crucially we also infer the initial conditions
$E(0), I^A(0), I^S(0)$.  Letting $S(0) = N-(E(0)+I^A(0)+I^S(0))$ and setting the rest of the parameters to zero, we have all initial conditions defined and the model may be solved numerically to obtain $\mu_D$ and $\mu_c$ to evaluate our likelihood.  We use the \textit{lsoda} solver available in the \textit{scipy.integrate.odeint} Python function.

To model interventions, a break point is established at which $\beta = \beta_1$ before and $\beta = \beta_2$ after the intervention day.  This creates a non-linear time dependent $\beta(t)$ \cite{wang2006,dehning2020inferring}.  This additional parameters are then included in the inference.  Normally only the initial $\beta_1$ and an after lockdown (22 March 2020) $\beta_2$ parameters are considered (in Mexico city a third $\beta_3$ was considered for modeling a further local intervention in early April).

We use $Gamma(1,10)$ priors for each $E(0), I^A(0), I^S(0)$, modeling the low, near to 10, and close to zero counts for the number of infectious initial conditions.  The prior for the $\beta$'s are the same and a priori independent (a possible generalization is to consider dependent $\beta$'s modeling a decrease in contact rates after lockdowns etc.).  These are rather diffuse
and long tails LogNormal with $\sigma^2 = 1$ and scale parameter 1; that is
$log(\beta) \sim N(0,1)$.

To sample from the posterior we resort to MCMC using the ``t-walk'' generic sampler \cite{christen2010twalk}.  The MCMC runs semi automatic, with a fairly consistent burn in of 1,000 iterations (sampling initial values from the prior).  We perform subsampling using the Intergrated Autocorrelation Time, with psuedo-independent sample sizes of 1,000 to 1,500 with 200,000 iterations of the MCMC.  This roughly takes 30 min in a 2.2 GHz processor.

To illustrate the whole posterior distribution, for any state variable $V$
(or $\mu_c(t_i)$), for each sampled initial conditions and $\beta$'s the model is solved at time $t_1, t_2, \ldots, t_k$, including possibly future dates, obtaining a sample of $V(t_i)$ values for each $t_i$.  The median, and other desired quantiles are plotted vertically for each date considered, obtaining the plots as in Figure~\ref{fig:cdmx}.  Note that the traced median or other plotted quantiles do not necessary correspond to any given model trajectory, providing a far richer Uncertainty Quantification approach than the classical 
parameter estimates plug-in approach.  Indeed, the sampled values for $V(t_i)$ do correspond to Monte Carlo samples of the posterior predictive distribution for $V(t_i)$.

\subsection*{Adding age structure}

Adding group ages is straightforward in these type of models.  The number $a$ of ages groups is decided and our model is repeated $a$ times.  Different residence times may be included \cite{moghadas2020} but we preferred to concentrate on the different transition probabilities $g, h$ that vary nearly two orders of magnitude using age groups $[0, 25], (25, 50], (50,65], (65, 100]$ \cite{ferguson2020impact}.  The age structure is used to divide the initial infectious and susceptible population proportional to the size of each age group.  The same number of parameters are inferred, using a single $\beta$, with an optional weighting contact matrix \cite{prem2017} to model different contact rates among age groups in specific regions. Using a sufficiently flexible software design, progressing to an age structure model is not complex, nonetheless the MCMC may run substantially slower.  Although we have experimented with our age structured model, using census data from Mexico and both uniform and non uniform contact matrices, we do not report any of these results here, given that our non-age structure model has sufficiently enough predictive power, as already discussed.

\subsection*{Confounding effect of $N_{eff}\times f$}

To explain the confounding effect of $N_{eff}\times f$
we have two observations.
First, if we let  $f= \tilde{f}/\alpha$ for some 
$\alpha\in (0,\tilde{f})$ then differential equations for the variables $I^c$, $H^1$, $H^2$, $H^3$, $U^1$, $U^2$ and $D$ remain invariant and  
the equations for $I^s$ becomes 
\begin{equation*}
\frac{dI^S}{dt}  =   
\frac{\Tilde{f}}{\alpha}\sigma_1 E - \sigma_2 I^S.
\end{equation*}
By letting $\Tilde{E}= E/\alpha$ the equation for $I^s$ 
is also invariant with the substitution of $E$ by $\Tilde{E}$. Now, the equation for $\Tilde{E}$ is given by 
\begin{equation*}
  \frac{d\Tilde{E}}{dt}  =  \frac{\left(\beta_{A} I^{A} +\beta_{S} I^{S} \right)}{ \alpha N_{eff}}  S - \sigma_1 \Tilde{E}. 
\end{equation*}
By letting $\Tilde{N}_{eff}=\alpha N_{eff}$ the latter equations becomes also invariant under the substitution of $E$ by $\Tilde{E}$. Therefore for 
the lower branch in the model (see Figure~\ref{fig:graph}) the system of equations is invariant under the change 
of $f$ and $N_{eff}$ by $\Tilde{f}$ and $\Tilde{N}_{eff}$ provided $\Tilde{N}_{eff}\times \Tilde{f}= N_{eff}\times f$ holds. 
Clearly, to have a consistent system of equations, the equations for $S$, $I^A$ and $R$ have to be adapted in each case to obtain a consistent set of equations.

Second, int the inference of parameter $\beta$ 
we inform the system with the data at the $H^1$ 
and $D$ compartments.  If $\Tilde{N}_{eff}\times \Tilde{f}= N_{eff}\times f$ holds, in view of our first observation, to fit these data the fluxes  $f\sigma_1 E$ and  $\tilde{f}\sigma_1 \tilde{E}$ in either case have to be the same. 
Clearly, the solutions in the compartment $I^S$ and after do
not change in this case, but the individuals in the $I^A$ 
compartment do change depending on which combination of  
$N_{eff}$ and $f$ or $\Tilde{N}_{eff}$ and $\Tilde{f}$ is considered. There is a range of 
validity for $\alpha$ where the inference of $\beta$
does not change but do not explore this property further. 

Numerical simulations where also preformed to confirm this confounding effect, but until the asymptomatic infection is 
fully described it is not possible to resolve this issue. 

\end{document}